\documentclass[a4paper,12pt]{article}
\usepackage{mathtext}
\usepackage[T2A]{fontenc}
\usepackage[utf8]{inputenc}
\usepackage[ukrainian,english]{babel}
\usepackage[dvips]{graphicx}

\def\nicefrac#1/#2{\leavevmode\kern.1em
\raise.5ex\hbox{\the\scriptfont0 #1}\kern-.1em
/\kern-.15em\lower.25ex\hbox{\the\scriptfont0 #2}}

\newcommand{\be}{\begin{equation}}
\newcommand{\ee}{\end{equation}}
\newcommand{\bea}{\begin{eqnarray}}
\newcommand{\eea}{\end{eqnarray}}
\newcommand{\bean}{\begin{eqnarray*}}
\newcommand{\eean}{\end{eqnarray*}}

\begin{document}

\title {Dynamics of expansion of the Universe in the models\\
with non-minimally coupled dark energy}
\author{R.\,Neomenko, \,\, B.\,Novosyadlyj\\
\it Astronomical Observatory of Ivan Franko National University of Lviv, \\
\it Kyryla i Methodia str., 8, Lviv, 79005, Ukraine}

\maketitle

\medskip

\begin{abstract}
We consider the dark energy model with barotropic equation of state, which interacts with dark matter through gravitation and another force, causing the energy-momentum exchange between them. Both components are described in approximation of ideal fluids, which are parametrized by density and equation of state parameters. Three types of interactions between dark components are considered: the interaction independent from their densities, the one proportional to density of dark energy and the one proportional to density of dark matter. The equations which describe the expansion dynamics of homogeneous and isotropic Universe and evolution of densities of both components for different values of interaction parameter are obtained on the bases of the general covariant conservation equations and Einstein's ones. For three kinds of interactions we show the existence of the range of values of parameters of dark energy for which the densities of dark components and their sum are negative. We 
find the conditions of positivity of density of dark energy and dark matter. The constraints on the value of parameter of interaction are derived. The dynamics of expansion of the Universe with these interactions of dark energy and dark matter is analysed.

\end{abstract}

\section*{Introduction}

One of the most popular explanations of accelerated expansion of the Universe is that it is caused by some scalar field, for which the strong condition of energy domination isn't satisfied: $\varepsilon_{de}+3p_{de}<0$. Different models of scalar field are considered: quintessence, phantom, quintom, tachyon and other more exotic models of scalar field (see for example, \cite{Copeland2006,Amendola2010,Novosyadlyj2013} and citations there in). To describe its impact on expansion of the Universe or other components the hydrodynamics approximation is used - the field represented by some ideal fluid with equation of state $p_{de}=w_{de}\rho_{de}$, where $w_{de}$ is called the parameter of equation of state. In the simplest case the quantity $w_{de}< -1/3$ is a constant, but there are more general models, where parameter of equation of state changes with time. Beside this, there are models, in which scalar field non-gravitationally interacts with other fields \cite{Amendola2000,Zimdahl2001,Gumjudpai2005,Campo2006,
Wei2007,Amendola2007,Guo2007,Caldera2009,Amendola2010}. In the  Lagrangian of scalar field and fields with which it interacts the additional term appears. In the general case this Lagrangian can be written in the form: $L=L_{\varphi}(X, \varphi)+L_{int}+L_{\psi}(\psi_{n}, \partial_{i}\psi_{n})$,
where $X=-\frac{1}{2}g^{ik}\partial_{i}\varphi\partial_{k}\varphi$ - kinetic term of field, $\varphi$ - the variable of scalar field, which is the dark energy, $\psi_{n}$ - the fields of other types of matter. In general, the form of $L_{int}$ is unknown. There are no any physical principle or any experimental data, from which it is possible to derive the Lagrangian of interaction of scalar field of dark energy with other fields yet. Nowadays, the attempts to build unified theory of field are being made, and the scalar field some how must be coupled with other fields. One can read about possibility of such interaction and observational data which points on it in works \cite{Amendola2003,Wei2007,Amendola2007,Guo2007,Vacca2009,Caldera2009,Abdalla2013,Penzo2015,Pollina2015,Elahi2015,Goncalves2015}.

 Below we analyse the impact of non-gravitational interaction between dark energy and dark matter on the dynamics of the expansion of the Universe and on the evolution of the densities of these components. As we don't know the form of the Lagrangian of scalar field we describe its behaviour phenomenologically. Evolution of homogeneous and isotropic Universe and its components are described by system of Einstein's equations and by equations which express the conservation laws of energy and momentum considering additional interaction between components. We assume that dark energy interacts gravitationally and non-gravitationally with dark matter only and interacts gravitationally only with baryonic matter or relativistic matter. All components are described by ideal fluid approximation. As the basis we took the model of dark energy \cite{Sergijenko2010,Novosyadlyj2010,Novosyadlyj2012,Novosyadlyj2013} which had been studied as minimally coupled, which gives the possibility to  distinguish the impact of 
interaction 
between dark components on dynamics of the expansion of the Universe and evolution of their densities.

\section{Einstein's equations and conservation laws for non minimally coupled components of the Universe}

The following features have been generally accepted as properties of the observational Universe: homogeneity, isotropy, flatness of 3-space and the accelerated expansion at large scales. The metric of 4-space of such world is the Friedmann-Lemaitre-Robertson-Walker metric (FLRW), which in comoving spherical coordinates in conformal representation have such form:
\begin{equation}\label{ds}
  ds^{2}=g_{ik}dx^idx^k=a^{2}(\eta)[d\eta^{2}-dr^{2}-r^{2}(d\theta^{2}+\sin^{2}\theta d\varphi^{2})],
\end{equation}
where $g_{ik}$ is the diagonal metric tensor with only one unknown function $a(\eta)$, the scale factor, which describes the expansion of the Universe. Hereafter indices $i, k, ...$ take the values 0, 1, 2, 3, where $x^0$-component will be always a time variable. In the model of Universe with zero curvature of 3-space it is convenient to normalize it by 1 at the present epoch: $a(\eta_0)=1$. The first and second time derivatives of this factor describe the rate and the acceleration of the expansion of the Universe, i.e. the dynamics of such expansion. The variable $\eta$ is conformal time, which is related with physical cosmological time $t$ by the simple differential relation: $cdt=a(\eta)d\eta$, $c$ - the speed of light. The large set of independent observational data (see the chapter 1 in the book \cite{Novosyadlyj2013}) indicates that $\frac{da}{dt}>0$ and $\frac{d^2a}{dt^2}>0$, or in conformal time $\frac{\dot{a}}{a}>0$ and $\frac{\ddot{a}}{a}-\left(\frac{\dot{a}}{a}\right)^2>0$, 
hereafter the dot denotes the derivative $\frac{d}{d\eta}$. Traditionally in cosmology the expansion rate of the Universe is described by relative quantity $H$, which is called the Hubble parameter, and the acceleration by the dimensionless quantity $q$, which is called the deceleration parameter:
$$H\equiv\frac{1}{a}\frac{da}{dt}=\frac{\dot{a}}{a^2}, \ \ \  q\equiv-\frac{1}{aH^2}\frac{d^2a}{dt^2}=-\frac{\ddot{a}}{a^3H^2}+1.$$
The equations for them are given by the Einstein's equations (1915), which in general case have the form
\begin{equation}\label{Rik}
  R_{ik}-\frac{1}{2}g_{ik}R=\frac{8\pi G}{c^4}T_{ik},
\end{equation}
where $R_{ik}$ is the Ricci tensor of covariant curvature, $R$ is its convolution, the scalar of 4-space curvature and $T_{ik}$ is the total covariant energy-momentum tensor of the components of the Universe. We specify the energy-momentum tensor in the following way: consider that the Universe is uniformly filled by non-relativistic matter - baryonic matter ($b$) and dark matter ($dm$), relativistic one ($r$) - cosmic microwave background and relict neutrino, and also dark energy ($de$). Each of it will be described by energy-momentum tensor of ideal fluid,
\begin{equation}\label{Tik}
  T_{i(N)}^{k}=(c^2\rho_{(N)}+p_{(N)})u_{i}u^{k}-p_{(N)}\delta_{i}^{k},
\end{equation}
where $\rho_{(N)}$ is the (energy) density of $N$-component, $p_{(N)}$ - its pressure, and $u_{i}$ - 4-vector of velocity. Then the Einstein's equations give the equations for $H$ and $q$
\begin{eqnarray}
  H^{2}&=&\frac{8\pi G}{3}\sum_N\rho_{(N)}, \label{H2} \\
  qH^2&=&\frac{4\pi G}{3}\sum_N(\rho_{(N)}+3p_{(N)}), \label{qH2}
\end{eqnarray}
called the Friedmann equations. Hereafter the variables, in which $c=1$, was used. To solve those equations one would need to specify the equations of state of each component and non-gravitational interaction between them, if such is present. Lets take the equation of state in the form $p_{(N)}=w_{(N)}\rho_{(N)}$, with the given known parameters of equation of state for non-relativistic and relativistic components ($w_{dm}=w_{b}=0$,  $w_{r}=1/3$) and unknown for dark energy $w_{dе}=w(a)$. We shall find it for scalar-field dark energy from the condition that the ratio of variations of pressure to density is constant and from specific form of non-gravitational interaction with dark matter.

To find the dependencies of densities on time $\eta$ (or on scale factor $a$) the equations (\ref{H2})-(\ref{qH2}) must be complemented by the conservation equations for each component. In general case, if there is any other interaction between components besides gravitational one, the equations of conservation of energy and momentum will be written in such form:
\begin{equation} 
  T_{i;k}^{k(N)}=J_{i}^{(N)}, \quad  \sum_{n}J_{i}^{(N)}=0,\label{JiN}
\end{equation}
where $J_{i}^{(N)}$ is the 4-vector of the change of density flow of energy-momentum to/from component $N$ as the result of interaction with other components, and ``;'' - denotes the covariant derivative with respect to $x^k$. For minimally coupled components $J_{i}^{(N)}=0$. The second equation in (\ref{JiN}) is the result of Bianchi identities. Hereafter we will suppose that non-gravitational interaction is present only between dark energy and dark matter, so the non-zeros are only $J_{i}^{(de)}$ - for dark energy and $J_{i}^{(dm)}$ - for dark matter. From the second equation in (\ref{JiN}) it follows that
\begin{equation}\label{Ji}
  J_{i}^{(de)}=-J_{i}^{(dm)}=J_{i}.
\end{equation}
Then the conservation laws for dark energy and dark matter will be written in following form:
\begin{equation}
  T_{i;k}^{k(de)}=J_{i}, \quad  T_{i;k}^{k(dm)}=-J_{i}. \label{tei_dedm}
\end{equation}
For baryonic matter and relativistic matter the conservation laws will be with zero flow of energy-momentum, caused by interaction, because we shall analyse  the dynamics of expansion of Universe in the epoch after recombination, when their motion is free and they interact only gravitationally:
\begin{equation}
  T_{i;k}^{k(b)}=0, \quad  T_{i;k}^{k(r)}=0. \label{tei_br}
\end{equation}
The conservation equations (\ref{tei_dedm}) and (\ref{tei_br}) for homogeneous isotropic Universe contain only the equations of discontinuity, which in the metrics (\ref{ds}) have the form
\begin{eqnarray}
  \dot{\rho}_{de}+3\frac{\dot{a}}{a}\rho_{de}(1+w)=J_{0}, \label{eq de} \\
  \dot{\rho}_{dm}+3\frac{\dot{a}}{a}\rho_{dm}=-J_{0}. \label{eq dm}
\end{eqnarray}
Since the Universe is homogeneous and isotropic on large scales and the perturbations are absent, the 4-vector of energy-momentum flow $J_{i}$ has only one non-zero component $J_{0}$ - the change of energy density per unit time.

Similarly to (\ref{eq de}) and (\ref{eq dm}) the equations for baryonic and relativistic components with zero right hand side give the well known expressions: $\rho_b(a)=\rho_b^{(0)}a^{-3}$, $\rho_r(a)=\rho_r^{(0)}a^{-4}$, hereafter up or down index (0)  will mean the quantity at the present epoch.

Now let us consider the dependence of parameter of equation of state $w$ on scale factor $a$. For this we shall introduce the new parameter - the adiabatic speed of sound $c_{a}^{2}=\dot{p}_{de}/\dot{\rho}_{de}$. Then, using equation (\ref{eq de}), we shall obtain such differential equation for $w(a)$:
\begin{equation}\label{dw}
  \frac{dw}{da}=\frac{3}{a}(1+w)(w-c_{a}^{2})-\frac{J_{0}}{\rho_{de}a^{2}H}(w-c_{a}^{2}).
\end{equation}
In general case the quantity $c_{a}^{2}$ can depend on time, which must be given or obtained from known or given other physical properties of dark energy. In this work we shall consider it as constant: $c_{a}^{2}=const\le0$ \cite{Novosyadlyj2010}. This equation must be solved together with equations (\ref{eq de}), (\ref{eq dm}), but for this the $J_{0}$ must be given or to derived from some considerations (see the reviews \cite{Copeland2006,Amendola2010,Bolotin2013}). As we don't know anything about such interaction, it is naturally to suggest that it is a function of energies of these two components:
\begin{equation}\label{JaHf}
  J_{0}=aHf(\rho_{de}, \rho_{dm}).
\end{equation}
For small values of densities it can be represented as
\begin{equation}\label{JaH}
  J_{0}=-3aH(\alpha+\beta\rho_{de}+\gamma\rho_{dm}),
\end{equation}
where $\alpha,\, \beta,\, \gamma$ - the constants, which give the strength and sign of interaction. We suppose, that interaction between dark energy and dark matter mustn't depend on the expansion rate of the Universe and conformal time $\eta$ explicitly. Therefore the expression for interaction $J_{0}$ we took in the form (\ref{JaHf}) and (\ref{JaH}) with factor $aH$, which removes the explicit dependence of interaction between dark components on the Hubble parameter $H$ and the conformal time $\eta$, as such factor is present in the right hand side of equations (\ref{eq de}), (\ref{eq dm}). This is obviously, if one switches from the differentiation with respect to $\eta$ to the differentiation with respect to $a$ in these equations:
$$\frac{\dot{a}}{a}\rightarrow aH, \quad \left(\dot{ }\right)\equiv\frac{d}{d\eta}\rightarrow a^2H\frac{d}{da}.$$
The change of dark components energy densities in this case is described by integral-differential equation, which can be solved numerically. In this article we consider only some partial cases of such interaction, for which the analytical solutions exist:
\begin{eqnarray}
  \beta=0, \gamma=0: && J_{0}=-3\alpha aH\rho_{cr}, \label{J0} \\
  \alpha=0, \gamma=0: && J_{0}=-3\beta aH\rho_{de}(a), \label{Jde} \\
  \alpha=0, \beta=0: && J_{0}=-3\gamma aH\rho_{dm}(a), \label{Jdm}
\end{eqnarray}
where $\rho_{cr}=3H_0^2/8\pi G$ is the critical density at present epoch. The component's densities at present epoch are convenient to represent in the units of critical one with the dimensionless parameter of density $\Omega_N$: $\rho_N^{(0)}=\Omega_N\rho_{cr}$.

In literature also the other kinds of interactions are considered, in particular, $J_{0}=Q\dot{\varphi}\rho_{dm}$, where $\varphi$ is the dark energy scalar field. \cite{Amendola2000,Pourtsidou2013,Penzo2015}. It can be rewritten in the form $J_{0}=-3\gamma(a) aH\rho_{dm}$, where $\gamma$ now depends on $a$. The models with such interaction and interaction (\ref{Jdm}) are considered in \cite{Amendola2010,Amendola2007}.

Further we shall consider the three cases of interaction (\ref{J0}), (\ref{Jde}) and (\ref{Jdm}) between dark energy and dark matter (the DE-DM interaction), which are reduced to analytical solutions of conservation equations (\ref{eq de}) and (\ref{eq dm}) for energy densities DE and DM and the equation (\ref{dw}) for equation of state parameter of dark energy.

Henceforth we shall call the dark energy as quintessence, if its density decreases in the process of expansion of Universe, and the dark energy, the density of which increases in the process of expansion of Universe - the phantom.

\section{DE-DM interaction independent from the densities of dark components}

\begin{figure}
\centering
\includegraphics[width=0.43\textwidth]{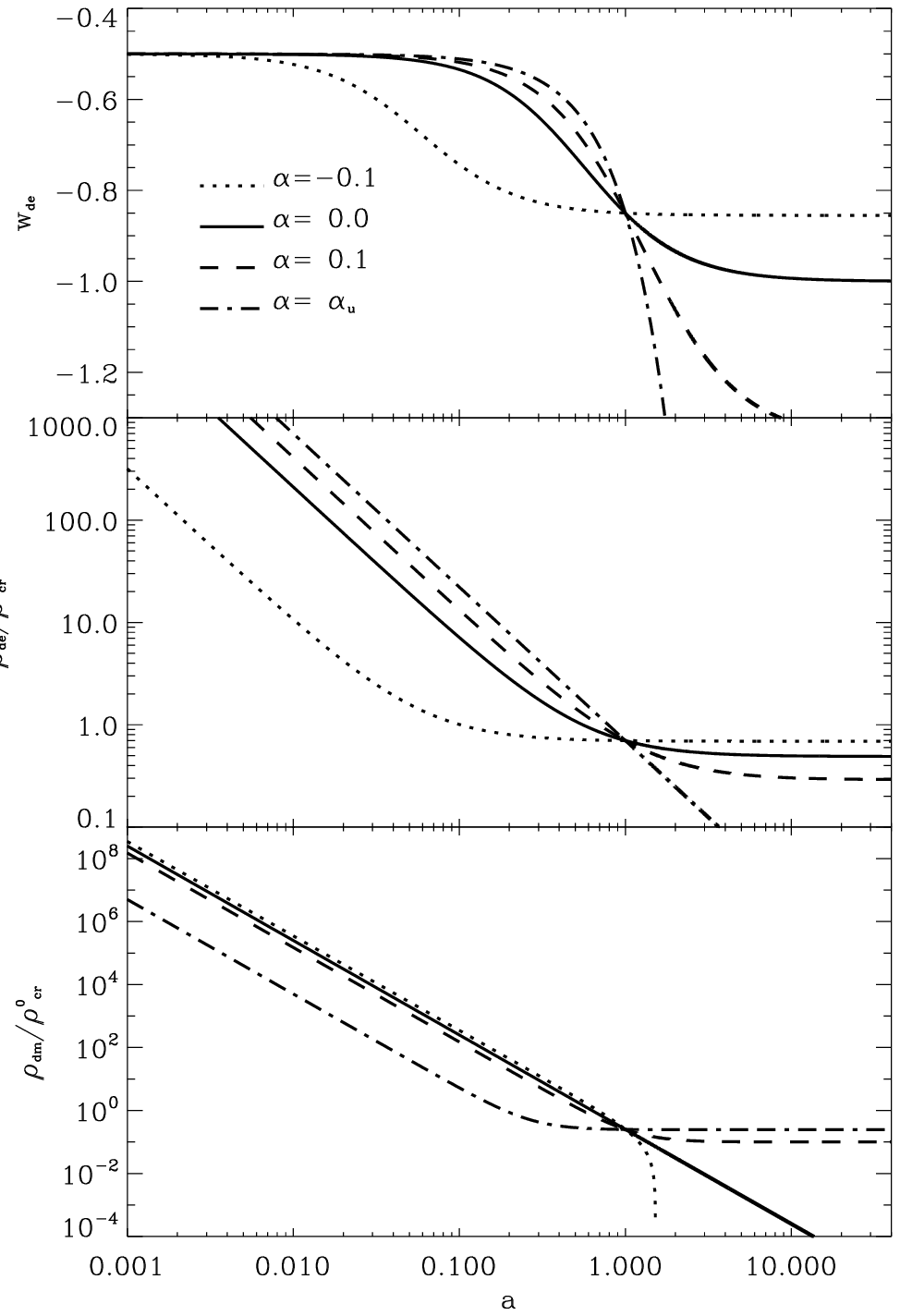}
\includegraphics[width=0.43\textwidth]{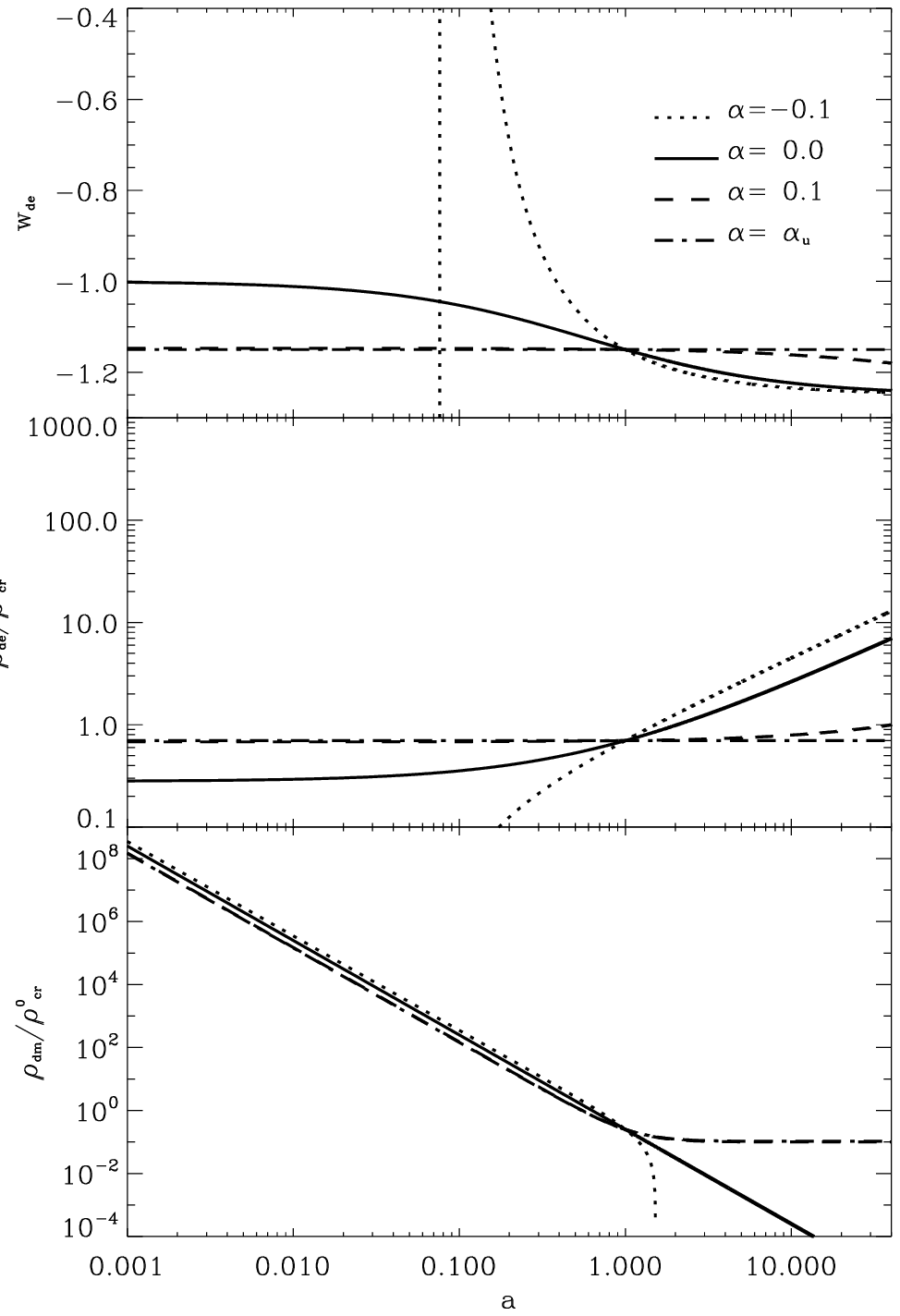}
\caption{The evolution of equation of state parameter of dark energy $w$, the density of dark energy $\rho_{de}$ and the density of dark matter $\rho_{dm}$ at the different values of interaction parameter $\alpha$ (the dotted line - $\alpha =-0.1$, the solid line - $\alpha =0$, the dashed line - $\alpha =0.1$, the dash-dotted line - $\alpha =\alpha_{u}$): a - the quintessence dark energy ($w_0=-0.85$, $c_a^2=-0.5$), b - the phantom ($w_0=-1.15$, $c_a^2=-1.25$). Here $\rho_{cr}^{0}$ is the critical density at present epoch ($a=1$), $\Omega_{de}=0.7$, $\Omega_{dm}=0.25$}
  \label{fig:rcr1}
\end{figure}

Consider the interaction (\ref{J0}), which does not depend on the densities of dark components. In this case from equations (\ref{eq de}) and (\ref{dw}) for arbitrary $J_0$ we obtain
\begin{equation}\label{rw}
  \rho_{de}=\rho_{de}^{(0)}\frac{w_0-c_a^2}{w-c_a^2},
\end{equation}
and as the result we have obtained such equation for $w$:
\begin{equation}\label{dwc}
  \frac{dw}{da}=\frac{3}{a}(w-c_{a}^{2})\left(1+w+\alpha\frac{\rho_{cr}}{\rho_{de}^{(0)}}
  \frac{w-c_{a}^{2}}{w_{0}-c_{a}^{2}}\right).
\end{equation}
This is the Riccati equation, which has the partial solution: $w=c_{a}^{2}$. With its help we find the general solution:
\begin{equation}\label{wc}
  w(a)=\frac{(1+c_{a}^{2})\left[(1+w_0)\Omega_{de}+\alpha(1-a^{3(1+c_{a}^{2})})\right]}{(1+w_0)\Omega_{de}-(w_0-c_a^2)\Omega_{de}a^{3(1+c_{a}^{2})}
  +\alpha(1-a^{3(1+c_{a}^{2})})}-1.
\end{equation}
One can see, that the character of equation of state parameter's evolution depends on the values of all parameters of dark energy and parameter of interaction $\alpha$. Moreover, if $\alpha\rightarrow0$, then $w$ goes to $w^{(mc)}$, the equation of state parameter of minimally coupled dark energy \cite{Novosyadlyj2010}:
$$w^{(mc)}(a)=\frac{(1+c_{a}^{2})(1+w_0)}{1+w_0-(w_0-c_a^2)a^{3(1+c_{a}^{2})}}-1.$$
Thus, the introduction of the even simplest interaction have changed the character of equation of state parameter's evolution significantly, which can be found from the comparison of fig. \ref{fig:rcr1} with fig. 1 in \cite{Novosyadlyj2010} and fig. 1 in \cite{Novosyadlyj2012}. In particular, the quintessence dark energy can gain in the future such properties, in which $w<-1$, while remaining the quintessence, namely such that density of which decreases in the process of expansion of the Universe. And vice versa for phantom. Considering (\ref{rw}) and (\ref{wc}) the conservation equations (\ref{eq de}) and (\ref{eq dm}) have exact analytical solutions:
\begin{eqnarray}
 \rho_{de}(a)&=&\rho_{de}^{(mc)}-\alpha\rho_{cr}\frac{1-a^{-3(1+c_a^2)}}{1+c_a^2},\label{r de0}\\
 \rho_{dm}(a)&=&\rho_{dm}^{(0)}a^{-3}+\alpha\rho_{cr}(1-a^{-3}), \label{r dm0}
\end{eqnarray}
where $\rho_{de}^{(mc)}$ is the known solution for minimally coupled dark energy ($\alpha=0$)
\begin{equation}\label{r demc}
  \rho_{de}^{(mc)}(a)=\rho_{de}^{(0)}\frac{(1+w_{0})a^{-3(1+c_{a}^{2})}-w_{0}+c_{a}^{2}}{1+c_{a}^{2}},
\end{equation}
which is smooth function of $a$ for the arbitrary parameters $w_0$ and $c_a^2$ \cite{Novosyadlyj2010}. The solutions (\ref{r de0}) and (\ref{r dm0}) are also the smooth functions for $0<a<\infty$ and arbitrary parameters of dark energy. They allow the negative values of density of dark components for certain values of parameters of dark energy and the parameter of interaction. The condition of positivity of density of dark matter for all $a$ is $0\leq\alpha\leq\Omega_{dm}$. The conditions of positivity of density of quintessence dark energy for all $a$ are $w_{0}<c_{a}^{2}, \alpha\le\Omega_{de}(c_a^2-w_0)$, and of phantom - $w_{0}<-1,\,\, \Omega_{de}(c_a^2-w_0)\leq\alpha\le-\Omega_{de}(1+w_0)$. Thereby, the interaction (\ref{J0}) provides the positive values of density of dark components only in the range of values of parameter of interaction
\begin{eqnarray}
&&0\le\alpha\le\min(\Omega_{dm}, \Omega_{de}(c_a^2-w_0)), \ \ \ \ w_{0}<c_{a}^{2}, \label{r demc1} \\
&&\max(0, \Omega_{de}(c_a^2-w_0))\le\alpha\le\min(\Omega_{dm}, -\Omega_{de}(1+w_0)), \ \ \ \ w_{0}<-1, \label{r demc2}
\end{eqnarray}
in the models with quintessence and phantom dark energy accordingly. It is interesting, that in the case of quintessence dark energy the expansion of Universe approaches exponential one and asymptotically constant energy densities of both dark components, and in the case of phantom - the singularity of Big Rip at constant values of density of dark matter.

Let us analyse the behaviour of the quantities $w$ and $\rho_{de}$ with the expansion of Universe considering the conditions (\ref{r demc1}) and (\ref{r demc2}). For the quintessence dark energy ($1+c_{a}^{2}>0$), when $a\rightarrow0$ we have $w\rightarrow c_{a}^{2}$, $\rho_{de}\rightarrow\infty$, and when $a\rightarrow\infty$ we have
$$w\rightarrow\frac{\alpha c_{a}^{2}-(w_{0}-c_{a}^{2})\Omega_{de}}{\alpha+(w_{0}-c_{a}^{2})\Omega_{de}},$$ $$\rho_{de}\rightarrow\frac{(c_{a}^{2}-w_{0})\Omega_{de}-\alpha}{1+c_{a}^{2}}\rho_{cr}.$$
For the phantom dark energy ($1+c_{a}^{2}<0$), when $a\rightarrow0$ we have
$$w\rightarrow\frac{\alpha c_{a}^{2}-(w_{0}-c_{a}^{2})\Omega_{de}}{\alpha+(w_{0}-c_{a}^{2})\Omega_{de}},$$ $$\rho_{de}\rightarrow\frac{(c_{a}^{2}-w_{0})\Omega_{de}-\alpha}{1+c_{a}^{2}}\rho_{cr},$$
and when $a\rightarrow\infty$ we have $w\rightarrow c_{a}^{2}$, $\rho_{de}\rightarrow\infty$. In the fig. \ref{fig:rcr1} the dependencies of quantities $w$, $\rho_{de}$ and $\rho_{dm}$ on $a$ in the range of its values $[10^{-3},\,40]$ are given. We see, that with values of parameter $\alpha$ at the upper bound for the model with quintessence dark energy ($\alpha_{u}=\Omega_{de}(c_a^2-w_0)$) when $a\rightarrow\infty$ we have $w\rightarrow-\infty$, $\rho_{de}\rightarrow0$. For the dark energy, with the parameters of phantom, for the upper bound ($\alpha_{u}=-\Omega_{de}(1+w_0)$) when $a\rightarrow\infty$ we have $w=const$, $\rho_{de}=const$, namely the dark energy in this case is similar to the cosmological constant. We see also that at negative $\alpha$ the density of dark matter $\rho_{dm}$ in the process of expansion of Universe at first is the positive, and then becomes negative.

\begin{figure}
\centering
\includegraphics[width=0.43\textwidth]{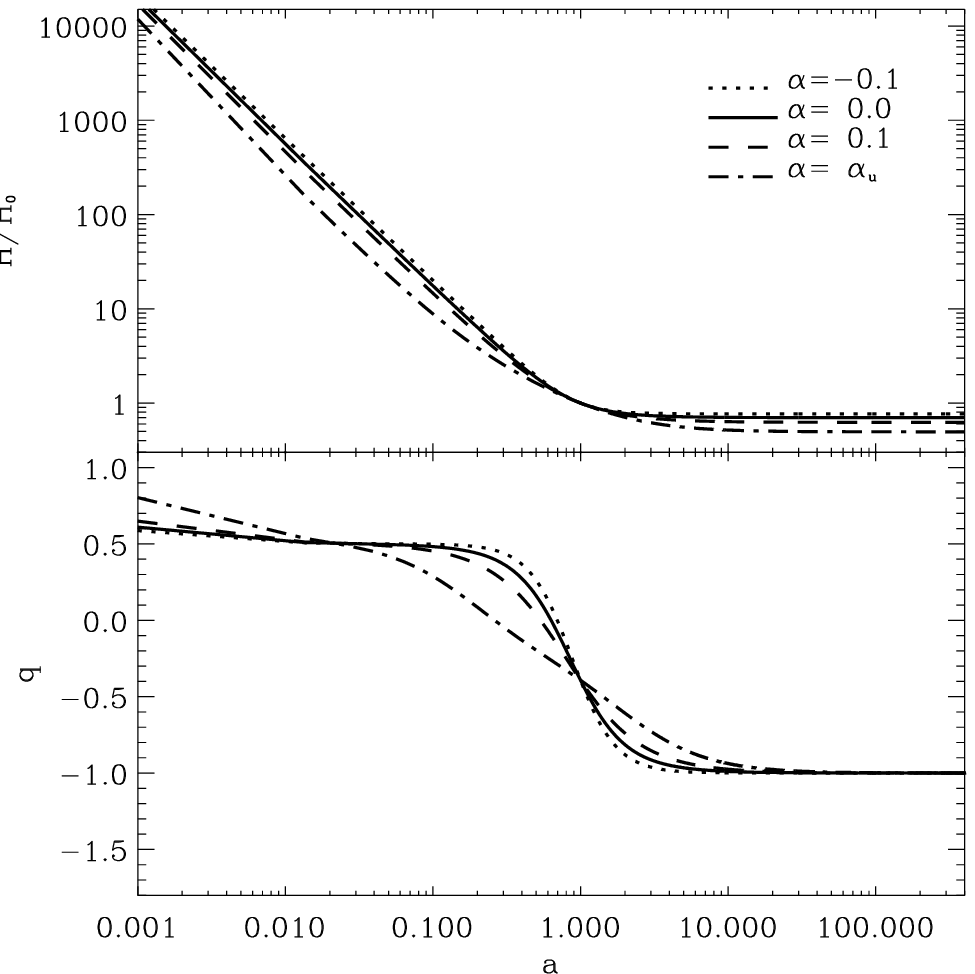}
\includegraphics[width=0.43\textwidth]{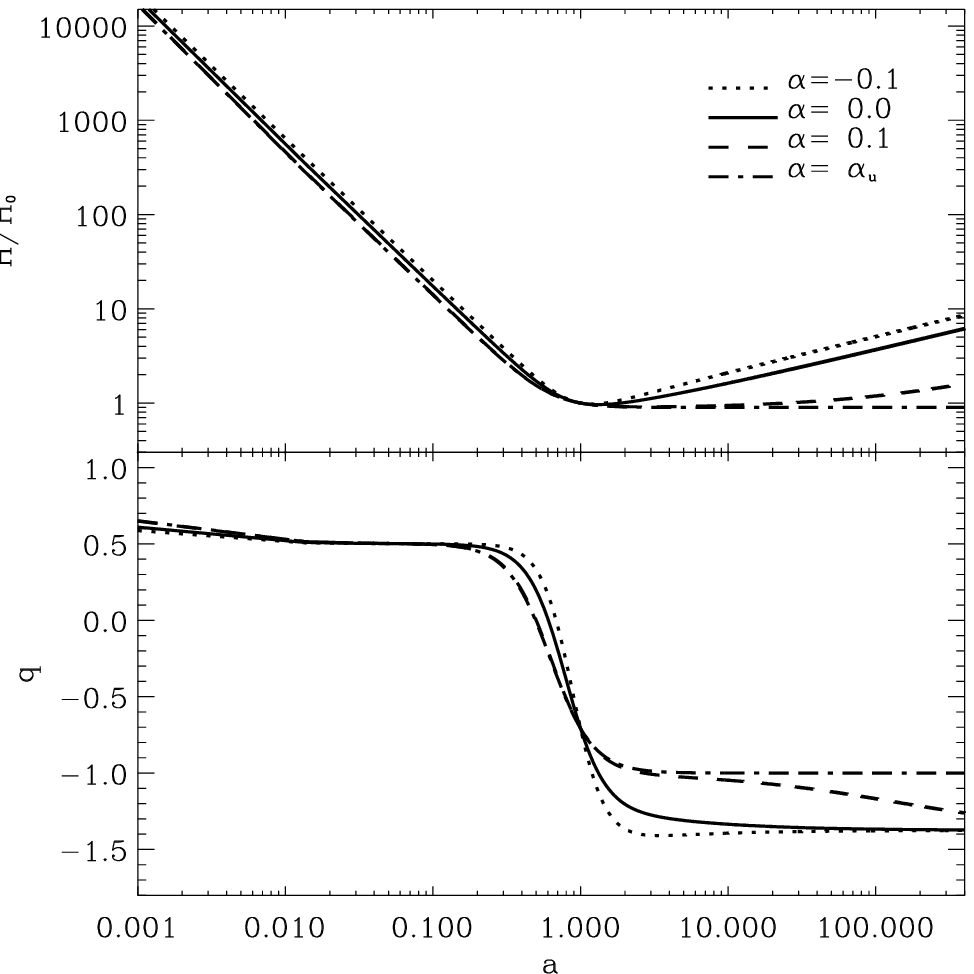}
\caption{The evolution of Hubble parameter $H$ and deceleration parameter $q$ at the different values of interaction parameter $\alpha$ (the dotted line - $\alpha =-0.1$, the solid line - $\alpha =0$, the dashed line - $\alpha =0.1$, the dash-dotted line - $\alpha =\alpha_{u}$): a - the model with quintessence dark energy, b - with the phantom. The parameters of models are the same as in the fig. \ref{fig:rcr1}}
\label{fig:rcr2}
\end{figure}

Let us consider now the impact of interaction (\ref{J0}) on the dynamics of expansion of Universe. For this we substitute in the Friedmann equations (\ref{H2}) and (\ref{qH2}) the  expressions for energy densities of dark energy (\ref{r de0}) and dark matter (\ref{r dm0})  founded above, and also the expressions for the density of baryonic matter $\rho_{b}=\rho_{b}^{(0)}a^{-3}$ and relativistic matter $\rho_{r}=\rho_{r}^{(0)}a^{-4}$. We calculate the values $H/H_0$ and $q$ for the standard cosmological model with $\Omega_{de}=0.7$, $\Omega_{dm}=0.25$, $\Omega_b=0.05$, $\Omega_r=4.17\cdot10^{-5}/h^2$ and $h=H_0/(100km\cdot s^{-1}Mpc^{-1})$. The results are shown in the fig. \ref{fig:rcr2}. The dependencies of $H/H_0$ and $q$ on $a$ are given for the range of its values $[10^{-3},\,400]$. Here at $\alpha=-0.1,\,0.0,\,+0.1$ for the quintessence dark energy, when $a\rightarrow0$, then $H/H_{0}\rightarrow\infty$, $q\rightarrow1$, and when $a\rightarrow\infty$, then
$$H/H_{0}\rightarrow\left(\frac{(c_{a}^{2}-w_{0})\Omega_{de}+\alpha c_{a}^{2}}{1+c_{a}^{2}}\right)^{1/2}, \ \ \ q\rightarrow-1.$$
For the phantom dark energy, when $a\rightarrow0$, then $H/H_{0}\rightarrow\infty$, $q\rightarrow1$, and when $a\rightarrow\infty$, then $H/H_{0}\rightarrow\infty$, $q\rightarrow\frac{1}{2}+\frac{3}{2}c_{a}^{2}$. For the upper bounds of interaction parameter $\alpha_{u}$ for the quintessence dark energy we have $H/H_{0}\rightarrow\infty$, $q\rightarrow1$ when $a\rightarrow0$ and $H/H_{0}\rightarrow(\Omega_{de}(c_{a}^{2}-w_{0}))^{1/2}$, $q\rightarrow-1$ when $a\rightarrow\infty$. For the dark energy with the upper bound $\alpha_{u}$ with the parameters of phantom we have the dark energy with constant $w$ and $\rho_{de}$. In this case $H/H_{0}\rightarrow\infty$, $q\rightarrow1$ when $a\rightarrow0$ and $H/H_{0}\rightarrow (-w_{0}\Omega_{de})^{1/2}$, $q\rightarrow-1$ when $a\rightarrow\infty$.

\section{DE-DM interaction proportional to the density of dark energy}

Let the expression for density of energy flow $J_{0}$ is such as in (\ref{Jde}). Then substituting it in the equation (\ref{dw}) for $w$, we shall obtain
\begin{equation}\label{dw1}
  \frac{dw}{da}=\frac{3}{a}(1+w)(w-c_{a}^{2})+\frac{3\beta}{a}(w-c_{a}^{2}).
\end{equation}
This equation has the exact analytical solution
\begin{equation}\label{w1}
w_{de}(a)=\frac{(1+c_a^2+\beta)(1+w_0+\beta)}{1+w_0+\beta-(w_0-c_a^2)a^{3(1+c_a^2+\beta)}}-1-\beta,
\end{equation}
where $w_0\equiv w(1)$ is the initial condition. Now, substituting the expression (\ref{w1}) for $w$ in expression (\ref{rw}) for density $\rho_{de}$, we obtain the exact analytical solution of equation (\ref{eq de}) for $\rho_{de}(a)$:
\begin{equation}\label{r de1}
  \rho_{de}(a)=\rho_{de}^{(0)}\frac{(1+w_{0}+\beta)a^{-3(1+c_{a}^{2}+\beta)}-w_{0}+c_{a}^{2}}{1+c_{a}^{2}+\beta}.
\end{equation}
As one can see, the dependencies $w(a)$ and $\rho_{de}(a)$ have three parameters $w_{0},\, c_{a}^{2},\, \beta$, which define the general properties and type of dark energy. Let us investigate the behaviour of $w$ and $\rho_{de}$, when $a\rightarrow0$ and $a\rightarrow\infty$. If $1+c_{a}^{2}+\beta>0$, then when $a\rightarrow0$ we have $w\rightarrow c_{a}^{2}$, $\rho_{de}\rightarrow\infty$, and when $a\rightarrow\infty$ we have $w\rightarrow-1-\beta$, $\rho_{de}\rightarrow\rho_{de}^{(0)}(c_{a}^{2}-w_{0})/(1+c_{a}^{2}+\beta)$. If $1+c_{a}^{2}+\beta<0$, then when $a\rightarrow0$ we have $w\rightarrow-1-\beta$, $\rho_{de}\rightarrow\rho_{de}^{(0)}(c_{a}^{2}-w_{0})/(1+c_{a}^{2}+\beta)$, and when $a\rightarrow\infty$ we have $w\rightarrow c_{a}^{2}$, $\rho_{de}\rightarrow\infty$. At the values of parameters $w_{0},\, c_{a}^{2}$ and $\beta$, for which the inequalities $w_0>c_a^2$ and $\beta<-(1+w_0)$, or $w_0<c_a^2$ and $\beta>-(1+w_0)$ are valid, the dependence of the equation of state parameter of dark energy $w$ 
on $a$ has the singularity of the second kind. From the definition of $w$ it is clear, that in that case the always smooth dependence $\rho_{de} (a)$ changes the sign in the process of expansion of Universe, namely $\rho_{de}(a_{\rho=0})=0$ at the point
$$a_{\rho=0}=\left(\frac{1+w_0+\beta}{w_0-c_a^2}\right)^{1/[3(1+c_a^2+\beta)]}.$$
Namely, the solution (\ref{r de1}) allows also $\rho_{de}<0$. We suppose, that the energy densities of the dark components of Universe can be only positive as during all its past and at the present epoch, and always in the future.

For the equation of state parameter and the density of dark energy we see here the new behaviour, different from the case $\beta=0$, studied in detail in the works \cite{Novosyadlyj2010,Novosyadlyj2012}.The phantom divide line is shifted by the quantity of interaction parameter $\beta$ and becomes equal $w_{phd}=-1-\beta$, if the quintessence dark energy is define as such, the density of which decreases ($1+c_{a}^{2}+\beta>0$), and the phantom dark energy is define as such, the density of which increases ($1+c_{a}^{2}+\beta<0$) in the process of expansion of the Universe.

The condition that $\rho_{de}\ge0$ at arbitrary $a\ge0$, for quintessence dark energy is the following:
\begin{equation}\label{um de1}
c_a^2\ge w_0, \quad \beta>-1-w_0,
\end{equation}
and for phantom one:
\begin{equation}\label{um de2}
c_a^2\le w_0, \quad \beta<-1-w_0.
\end{equation}

Now we shall find the dependence of density of dark matter $\rho_{dm}$ on $a$. Substituting the expression (\ref{r de1}) in the equation (\ref{eq dm}), we obtain
\begin{equation}\label{eq1 dm}
  \frac{d\rho_{dm}}{da}+\frac{3}{a}\rho_{dm}=\frac{3}{a}\beta\rho_{de}^{(0)}\displaystyle\biggl(Aa^{-3(1+c_{a}^{2}+\beta)}-B\biggr),
\end{equation}
where
\begin{equation}\label{AB}
  A=\frac{1+w_{0}+\beta}{1+c_{a}^{2}+\beta}, \quad B=\frac{w_{0}-c_{a}^{2}}{1+c_{a}^{2}+\beta}.
\end{equation}
Note, that the right hand side of equation (\ref{eq1 dm}) there is a regular function for an arbitrary finite value $0<a<\infty$ and arbitrary values of parameters $w_0$, $c_a^2$ and $\beta$. In the case of $1+c_a^2+\beta=0$, when $\rho_{de}=\rho_{de}^{(0)}=const$ and $w=w_0=const$, the solution of (\ref{eq1 dm}) is
\begin{equation}
\rho_{dm}(a)=\left(\rho_{dm}^{(0)}-\beta\rho_{de}^{(0)}\right)a^{-3}+\beta\rho_{de}^{(0)}. \nonumber
\end{equation}
The condition that $\rho_{dm}\ge0$ for any $0<a<\infty$, is the restriction of the range of values of the interaction parameter:
\begin{equation}
0\le \beta\le \frac{\Omega_{dm}}{\Omega_{de}}.\label{um dm11}
\end{equation}
So, in this particular case $\rho_{de}=const$ the interaction between the dark matter and dark energy of the form (\ref{Jde}) is such, that the energy flows from the dark energy to dark matter. The flow rate decreases in the process of expansion of the Universe and goes to an asymptotic regime $J_0\propto-\beta a H_0\rho_{de}^{(0)}$ such that $\rho_{dm}\rightarrow \beta\rho_{de}^{(0)}$ when $a\rightarrow\infty$.
\begin{figure}
\centering
\includegraphics[width=0.43\textwidth]{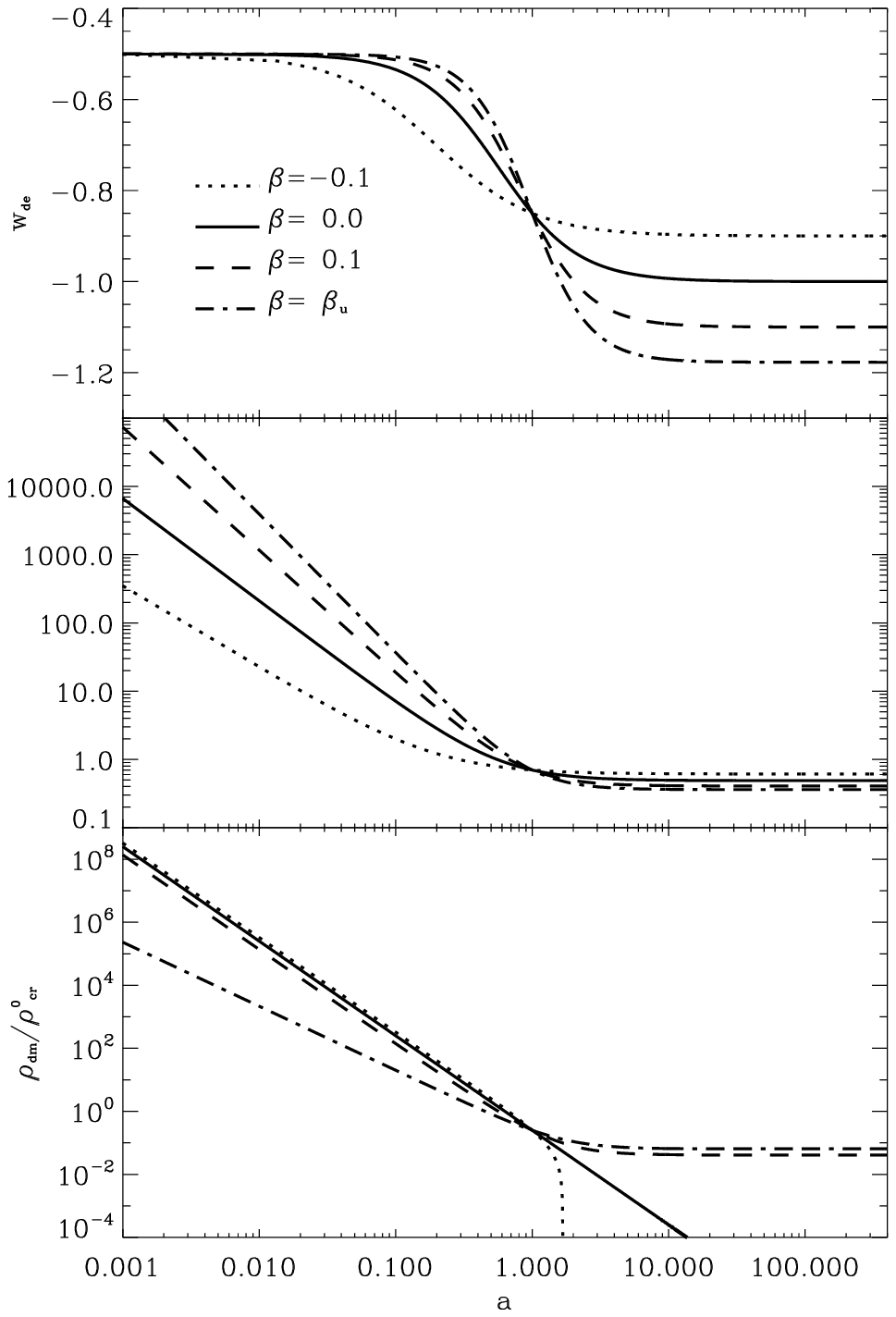}
\includegraphics[width=0.43\textwidth]{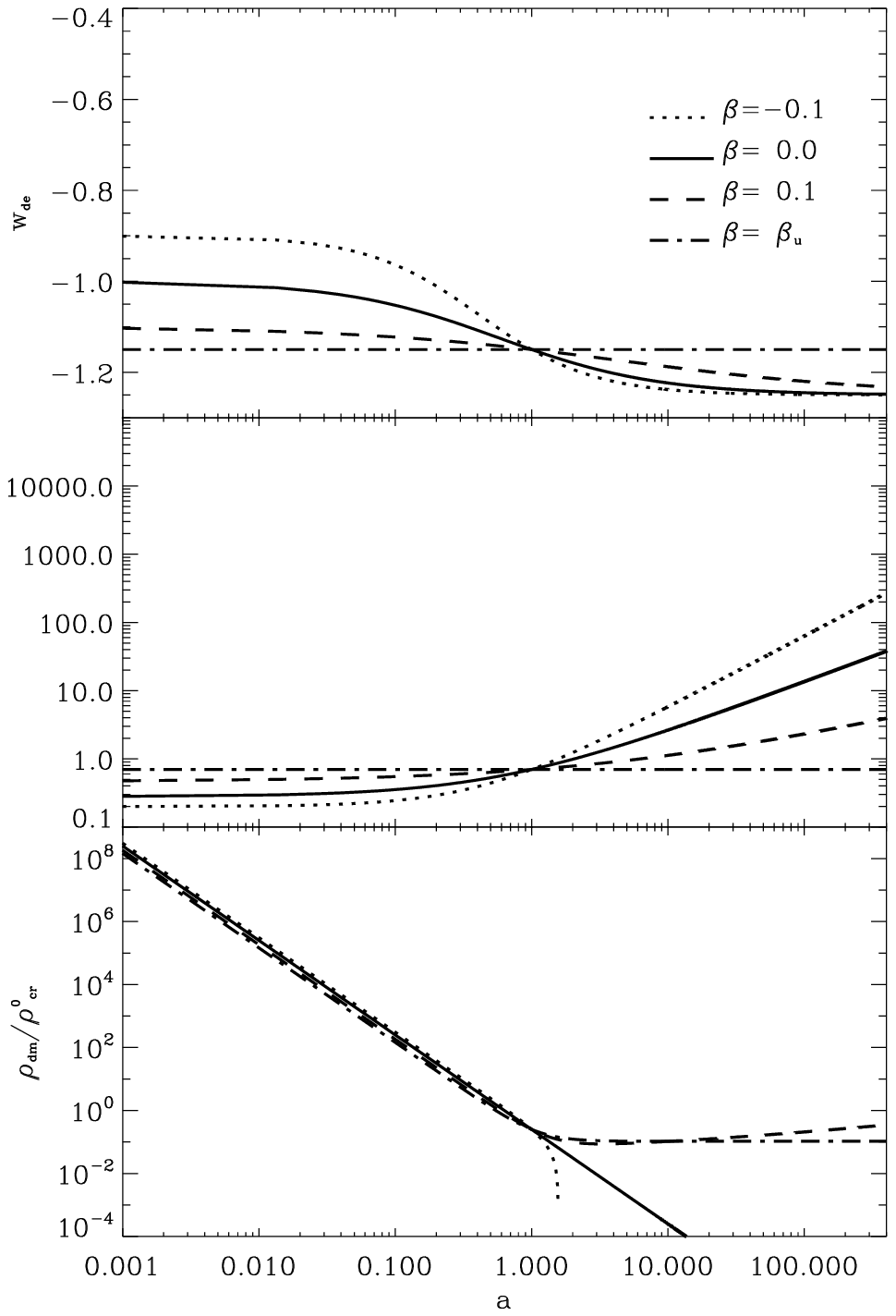}
\caption{The evolution of equation of state parameter of dark energy $w$, the density of dark energy $\rho_{de}$ and the density of dark matter $\rho_{dm}$ at the different values of interaction parameter $\beta$ (the dotted line - $\beta =-0.1$, the solid line - $\beta =0$, the dashed line - $\beta =0.1$, the dash-dotted line - $\beta =\beta_{u}$): a - the quintessence dark energy ($w_0=-0.85$, $c_a^2=-0.5$), b - the phantom one ($w_0=-1.15$, $c_a^2=-1.25$). Here $\rho_{cr}^{0}$ is the critical density at present epoch ($a=1$), $\Omega_{de}=0.7$, $\Omega_{dm}=0.25$}
  \label{fig:de1}
\end{figure}

The general solution of equation (\ref{eq1 dm}) is
\begin{equation}
\rho_{dm}(a)=\rho_{dm}^{(0)}a^{-3}+\beta\rho_{de}^{(0)}\displaystyle\biggl[
\biggl(\frac{A}{c_{a}^{2}+\beta}+B\biggr)a^{-3}-\frac{A}{c_{a}^{2}+\beta}a^{-3(1+c_{a}^{2}+\beta)}-B\biggr]. \label{r dm1}
\end{equation}
One can see, that the solution is regular at the all interval $0<a<\infty$ for an arbitrary values of parameters $w_0$, $c_a^2$ and $\beta$. If $w_{0}=c_{a}^{2}$, then $w=const$, and the expressions for energy densities of dark components (\ref{r de1}) and (\ref{r dm1}) coincide with the corresponding expressions in the work \cite{Abdalla2013}. In the particular case $c_a^2+\beta=0$, which is the case for quintessence dark energy,
$$\rho_{de}(a)=(1+w_0+\beta)\rho_{de}^{(0)}a^{-3}-(w_0+\beta)\rho_{de}^{(0)},$$
and the expression (\ref{r dm1}) is simplified:
\begin{equation}
\rho_{dm}(a)=\left(\rho_{dm}^{(0)}+\beta(\beta+w_0)\rho_{de}^{(0)}\right)a^{-3}-\beta(\beta+w_0)\rho_{de}^{(0)}. \nonumber
\end{equation}
It is easy to see, that in this case the condition $\rho_{dm}\ge0$ for an arbitrary $0<a<\infty$ together with the condition $\rho_{de}\ge0$ (\ref{um de1}) is satisfied at
\begin{equation}
  \max(0, -1-w_{0})\le \beta\le -w_0, \label{um dm22}
\end{equation}
and at the realistic values of equation of state parameter $w_{0}^{2}<4\Omega_{dm}/\Omega_{de}$ \cite{Sergijenko2015}.

Let us find the range of values $\beta$, at which $\rho_{dm}\ge0$ for an arbitrary $0<a<\infty$ and an arbitrary parameters of dark energy. In the case of quintessence dark energy with $\rho_{de}\ge0$ the value of density of dark matter $\rho_{dm}$ always will be positive with the condition
\begin{equation}\label{um dm0}
\max(0, -1-w_0)\le\beta\le-\frac{c_a^2\Omega_{dm}/\Omega_{de}}{1+w_0-c_a^2+\Omega_{dm}/\Omega_{de}}, \ \ \ \ \beta\neq-1-w_0,
\end{equation}
and in the case of phantom - with condition
\begin{eqnarray}
0\le\beta\le\min\left(-1-w_0,-\frac{c_a^2\Omega_{dm}/\Omega_{de}}{1+w_0-c_a^2+\Omega_{dm}/\Omega_{de}}\right), \ \ \ \ \beta\neq-1-w_0.   \label{um dm1}
\end{eqnarray}
Here the conditions of positivity of density of dark energy (\ref{um de1}) and (\ref{um de2}) are take into account. The upper bound of values of $\beta$ from the inequalities (\ref{um dm0}) and (\ref{um dm1}) we denote $\beta_u$.

On fig. \ref{fig:de1} we represent the dependencies $w(a)$, $\rho_{de}(a)$ and $\rho_{dm}(a)$ for the model of Universe with quintessence (left) and phantom (right) dark energy and the three values of interaction parameter $\beta=-0.1,\,0.0,\,+0.1$, and its upper limit value $\beta=\beta_u$. The parameters are chosen the same as in the fig. \ref{fig:rcr1}. The impact of value of interaction parameter $\beta$ on the evolution of equation of state parameter is equivalent to the shift of phantom divide line ($w=-1$) on the quantity $\beta$. In the case of $\beta>0$, when the energy flows from dark energy to dark matter, the density of quintessence dark energy decreases faster, and the density of phantom one increases slower, than in the case without of interaction ($\beta=0$). In the case of quintessence dark energy the density of dark matter goes to the constant value, and in the case of phantom - slowly goes to infinity. Also we see, that for the upper limit value of interaction parameter $\beta_u=-1-w_{0}$ 
for the dark 
energy with parameters of phantom we have the case, which we have considered earlier, in which the quantities $w$ and $\rho_{de}$ are constant. In the case of $\beta<0$, when the energy flows from dark matter to dark energy, the density of quintessence dark energy decreases slower, and of phantom increases faster, than in the case without the interaction. The density of dark matter in this case decreases fast, and goes to zero and to the negative values in future, which we consider as non-physical behaviour. This means, that the interaction of type (\ref{Jde}) can take place only when $\beta>0$.

\begin{figure}
\centering
\includegraphics[width=0.43\textwidth]{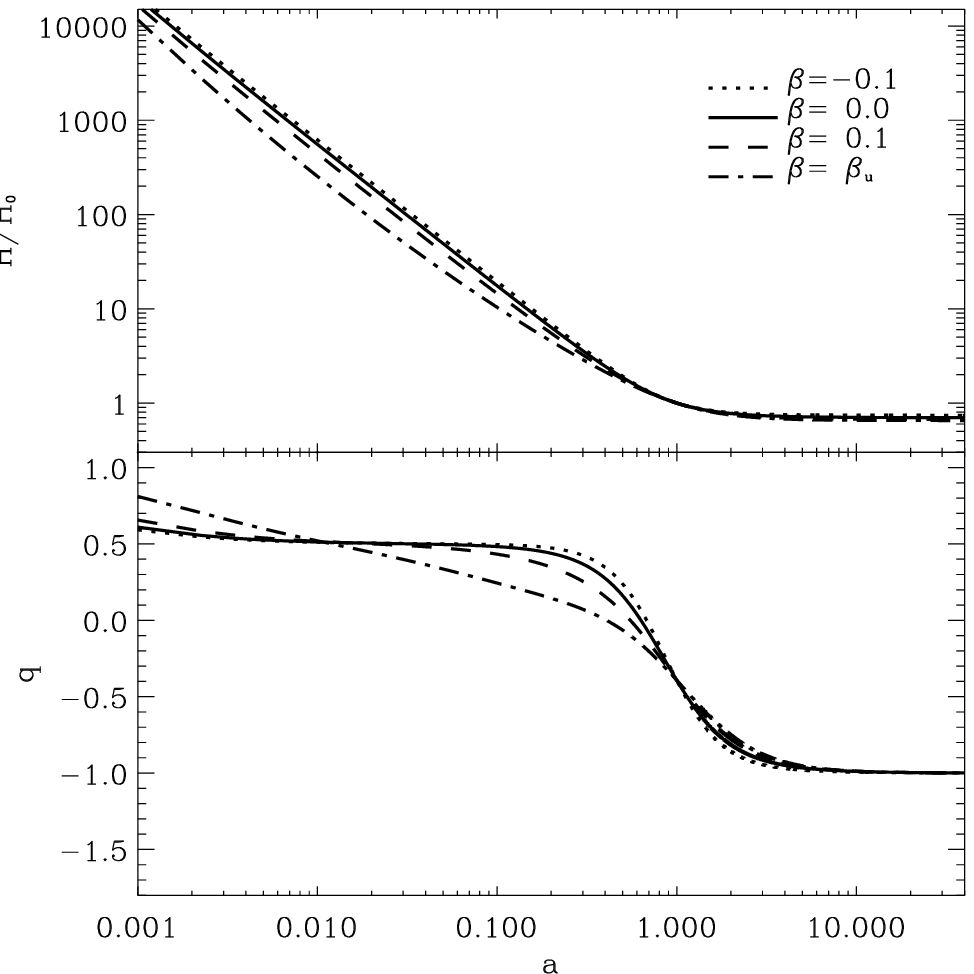}
\includegraphics[width=0.43\textwidth]{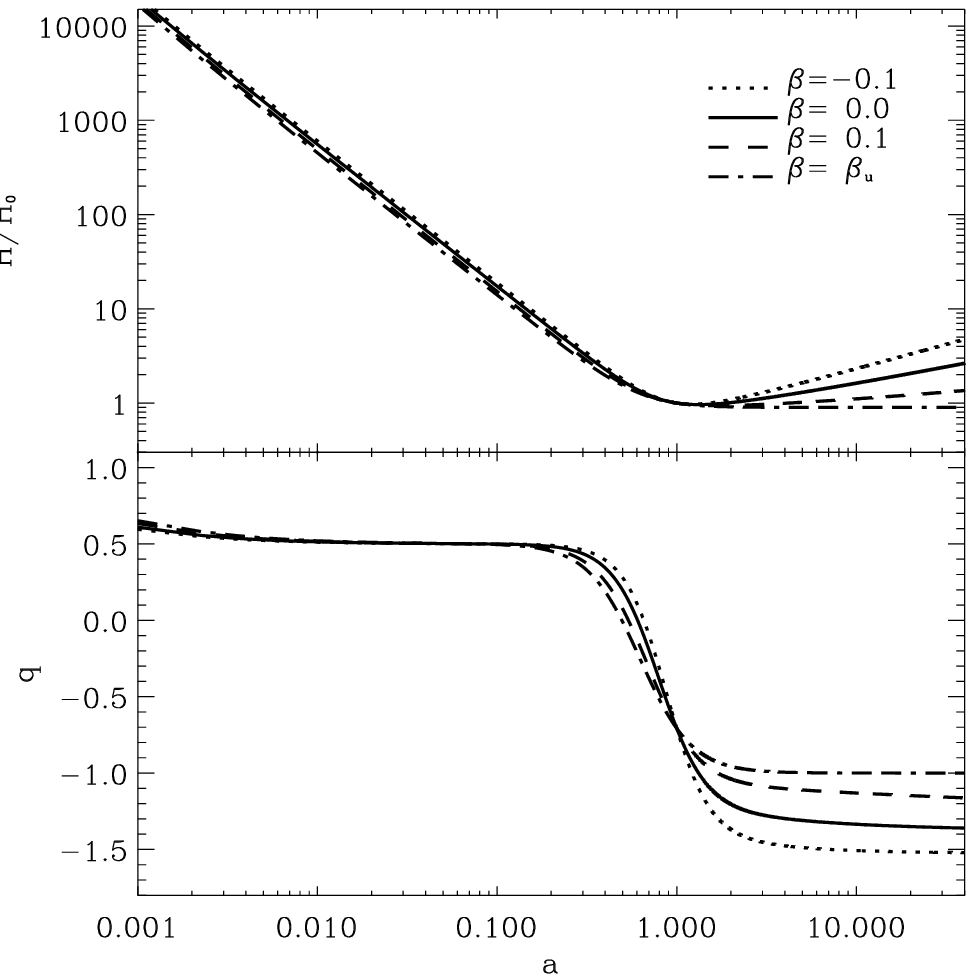}
\caption{The evolution of Hubble parameter $H$ and deceleration parameter $q$ at the different values of interaction parameter $\beta$ (the dotted line - $\beta =-0.1$, the solid line - $\beta =0$, the dashed line - $\beta =0.1$, the dash-dotted line - $\beta =\beta_{u}$): a - the quintessence dark energy, b - the phantom one. The parameters of models are the same as in the fig. \ref{fig:rcr1}}
\label{fig:de2}
\end{figure}
Consider now the impact of interaction (\ref{Jde}) on the dynamics of expansion of Universe. Substituting the expressions for density of all components in the Friedmann equations (\ref{H2}) and (\ref{qH2}), we obtain the dependence of quantities $H/H_{0}$ and $q$ on $a$. The results are represented in the fig. \ref{fig:de2} for the same parameters as in the fig. \ref{fig:rcr2}: left - the model with quintessence dark energy, right - with phantom one. On the all panels of both figures there are also the curves, which correspond to the upper bounds of values $\beta$ from inequalities (\ref{um dm0}) and (\ref{um dm1}). Here we see, that for the quintessence dark energy ($1+c_{a}^{2}+\beta>0$) $H/H_{0}\rightarrow\infty$, $q\rightarrow1$ when $a\rightarrow0$ and $H/H_{0}\rightarrow-\Omega_{de}B(1+\beta)$, $q\rightarrow-1$ when $a\rightarrow\infty$. For the phantom ($1+c_{a}^{2}+\beta<0$) $H/H_{0}\rightarrow\infty$, $q\rightarrow1$ when $a\rightarrow0$ and $H/H_{0}\rightarrow\infty$, $q\rightarrow \frac{1}{2}
+\frac{3}{2}(c_{a}^{2}+\beta)$ when $a\rightarrow\infty$. For the upper bound of interaction parameter ($\beta_u=-1-w_{0}$), for the dark energy with constant $w$ and $\rho_{de}$ with parameters of phantom $H/H_{0}\rightarrow\infty$, $q\rightarrow1$ when $a\rightarrow0$ and $H/H_{0}\rightarrow (-w_{0}\Omega_{de})^{1/2}$, $q\rightarrow-1$ when $a\rightarrow\infty$. As we see, such interaction with the considered values of parameter $\beta$ has impact on the dynamics of expansion of Universe in the past and present epochs (on figure $a\le1$), but this impact is not large. This means, that the high precision data is needed for finding the constraints of quantity $\beta$ .

\section{DE-DM interaction proportional to the density of dark matter}

Let now the expression for energy density flow $J_{0}$ is such as in (\ref{Jdm}). Substituting it in the equation (\ref{eq dm}) for $\rho_{dm}$ we obtain
\begin{equation}\label{eq2 dm}
  \frac{d\rho_{dm}}{da}+\frac{3}{a}\rho_{dm}=\frac{3\gamma}{a}\rho_{dm}.
\end{equation}
The general solution of this equation is the expression
\begin{equation}\label{r dm2}
  \rho_{dm}(a)=\rho_{dm}^{(0)}a^{-3(1-\gamma)}.
\end{equation}
So, the density of dark matter is always smooth function, which takes only the positive values for an arbitrary $0<a<\infty$ and $\gamma$. In the case of $\gamma>0$ (the flow of energy from dark energy to dark matter) the density of dark matter decreases slower for $\gamma<1$, than in the case of non-interacting components. The observational data on the large-scale structure of the Universe and anisotropy of cosmic microwave background constrain it value, $|\gamma|\ll1$.

Using the formula (\ref{rw}), we obtain the following ordinary differential equation for $w$:
\begin{equation}\label{dw2}
\frac{dw}{da}=\frac{3}{a}(w-c_{a}^{2})\left(1+w+\gamma\frac{\Omega_{dm}a^{-3(1-\gamma)}}{\Omega_{de}}
\frac{w-c_{a}^{2}}{w_{0}-c_{a}^{2}}\right).
\end{equation}

As in the previous cases, we have the Riccati equation with partial solution $w=c_{a}^{2}$, using which the general solution can be found easily:
\begin{equation}\label{w2}
  w(a)=\frac{\left[1+w_0+\gamma(1+c_{a}^{2})\frac{\Omega_{dm}}{\Omega_{de}}\frac{1-a^{3(c_a^2+\gamma)}}{c_a^2+\gamma}\right](1+c_{a}^{2})}
  {1+w_0+\gamma(1+c_{a}^{2})\frac{\Omega_{dm}}{\Omega_{de}}\frac{1-a^{3(c_a^2+\gamma)}}{c_a^2+\gamma}-(w_0-c_a^2)a^{3(1+c_a^2)}}-1
\end{equation}
We substitute it in the expression for $\rho_{de}$ (\ref{rw}) and obtain
\begin{equation}\label{r de2}
  \rho_{de}(a)=\rho_{de}^{(0)}\left[\frac{(1+w_{0})a^{-3(1+c_{a}^{2})}+c_{a}^{2}-w_{0}}{1+c_{a}^{2}}+
  \gamma\frac{\Omega_{dm}}{\Omega_{de}}\frac{1-a^{3(c_a^2+\gamma)}}{c_{a}^{2}+\gamma}a^{-3(1+c_{a}^{2})} \right].
\end{equation}

First of all let us notice, that the obtained solutions for the density of dark matter $\rho_{dm}(a)$ and dark energy $\rho_{de}(a)$ are the regular functions for all range of values of scale factor $0<a<\infty$ for an arbitrary values of parameters of dark energy $\Omega_{de}$, $w_0$, $c_a^2$ and interaction parameter $\gamma$. Indeed, the expression (\ref{r de2}) is finite when $1+c_a^2\rightarrow0$, and when $c_a^2+\gamma\rightarrow0$. In the particular case of constant $w$, when $w_{0}=c_{a}^{2}$, the expression (\ref{r de2}) coincide with the expression for the density of dark energy, given in the work \cite{Amendola2007}.

Let us analyse the behaviour of $w(a)$ and $\rho_{de}(a)$ when $a\rightarrow0$ and $a\rightarrow\infty$ considering the condition $|\gamma|\ll1$. If $1+c_{a}^{2}>0$, then for $a\rightarrow0$ we have $w\rightarrow c_{a}^{2}$, $\rho_{de}\rightarrow\infty$, and for $a\rightarrow\infty$ we have $w\rightarrow-1$, $\rho_{de}\rightarrow-\rho_{de}^{(0)}\frac{w_{0}-c_{a}^{2}}{1+c_{a}^{2}}$ and if $c_a^2\ge w_0$ this asymptotic value of the density of dark energy is positive. While if $1+c_{a}^{2}<0$, than when $a\rightarrow0$ we have $w\rightarrow c_{a}^{2}$, $\rho_{de}\rightarrow\infty$, and when $a\rightarrow\infty$ we have $w\rightarrow c_{a}^{2}$, $\rho_{de}\rightarrow\infty$. So, as before, when $1+c_{a}^{2}>0$ we have the quintessence dark energy, and when $1+c_{a}^{2}<0$ - the phantom one.

\begin{figure}
\centering
\includegraphics[width=0.43\textwidth]{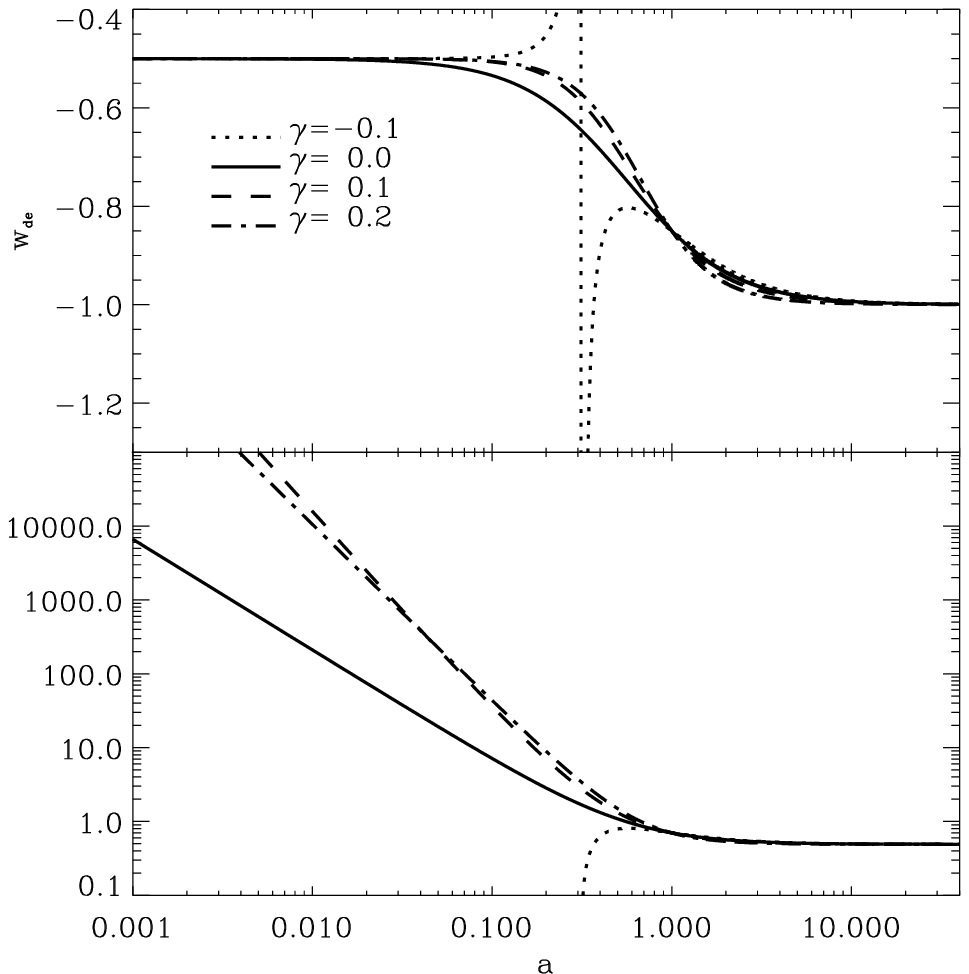}
\includegraphics[width=0.43\textwidth]{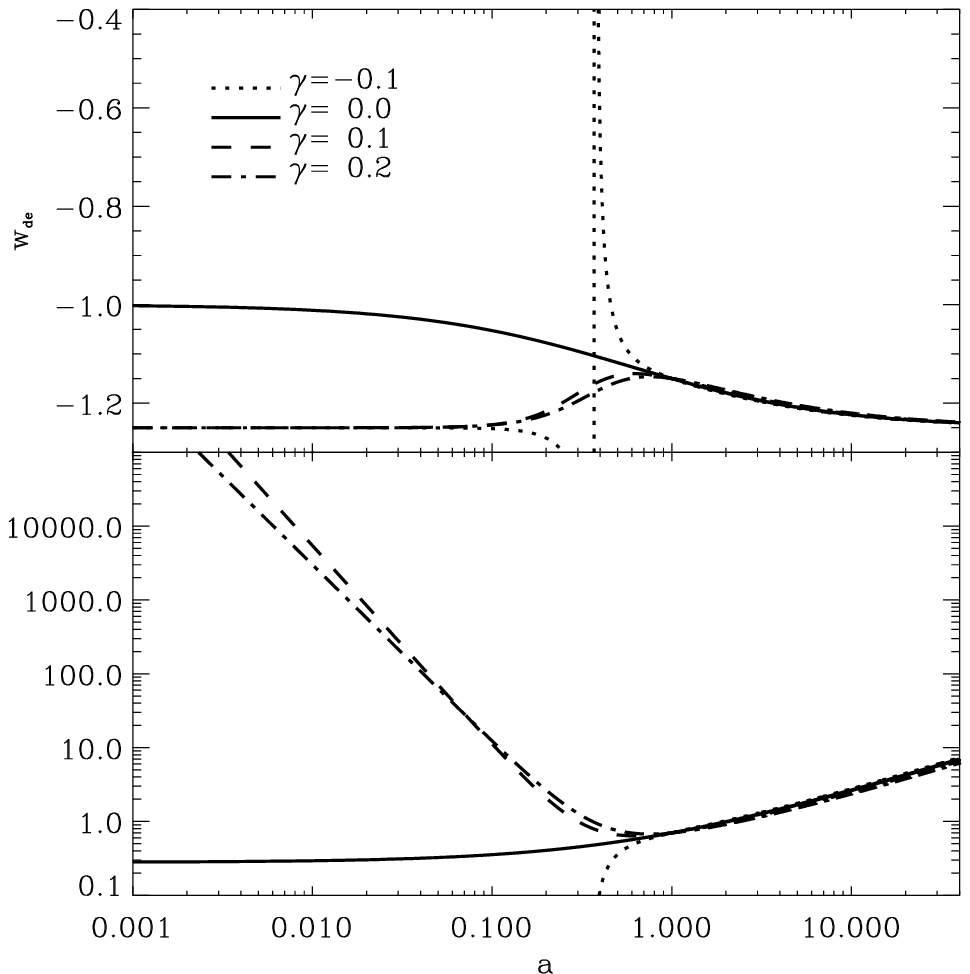}
\caption{The evolution of equation of state parameter of dark energy $w$, the density of dark energy $\rho_{de}$ at the different values of interaction parameter $\gamma$ (the dotted line - $\gamma =-0.1$, the solid line - $\gamma =0$, the dashed line - $\gamma =0.1$, the dash-dotted line - $\gamma =0.2$): a - the quintessence dark energy, b - the phantom. The parameters are the same as in the fig. \ref{fig:rcr1}}
  \label{fig:dm1}
\end{figure}

Let us write now the conditions of positivity for the density of dark energy $\rho_{de}$ considering the condition $\gamma\ll1$ for the realistic values of parameters $w_{0}$, $c_{a}^{2}$, $\Omega_{de}$, $\Omega_{dm}$. The density $\rho_{de}$ is always positive, if and only if $\gamma\geq0$ and $c_{a}^{2}+\gamma<0$. For the quintessence dark energy with $w_{0}>-1$ the conditions are following:

\begin{eqnarray}
  w_{0}\leq c_{a}^{2}, \ \ \ \ 0\leq\gamma\ll1.  \label{um2 de1}
\end{eqnarray}
For the phantom dark energy when $w_{0}<-1$ the conditions are following:
\begin{eqnarray}
  w_{0}\geq c_{a}^{2},\ \ \ \ 0\leq\gamma\ll1, \label{um2 de2} \\
  w_{0}<c_{a}^{2},\ \ \ \ 0<\gamma\ll1. \label{um2 de3}
\end{eqnarray}

From the conditions (\ref{um2 de2}), (\ref{um2 de3}) one can see, that when $\gamma=0$ the phantom dark energy can be the always positive only if $w_{0}\geq c_{a}^{2}$.
From the formula (\ref{w2}) one can see also, that there are such models, for which $w$ may cross the line $-1$, when the quantity $\rho_{de}$ is always positive. We do not give here the conditions of positivity of density of dark energy for them because they are bulky and the parameter values region corresponding to them is narrow.

\begin{figure}
\centering
\includegraphics[width=0.43\textwidth]{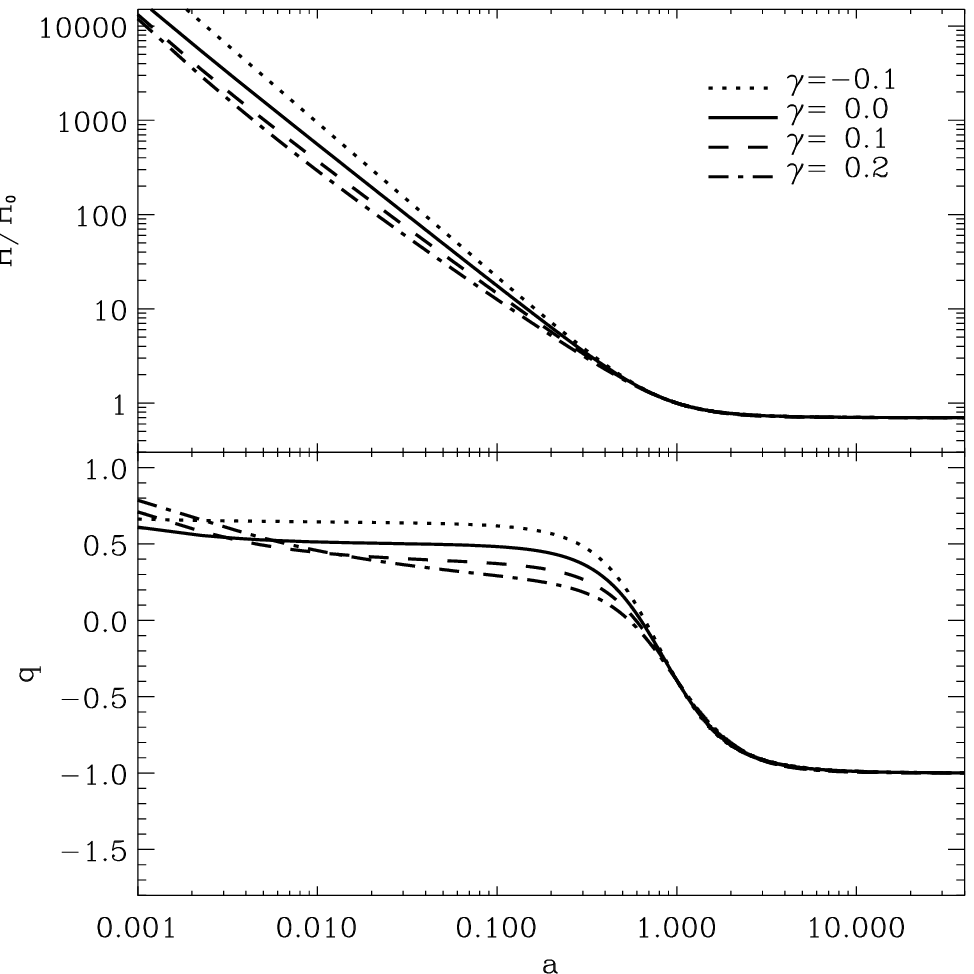}
\includegraphics[width=0.43\textwidth]{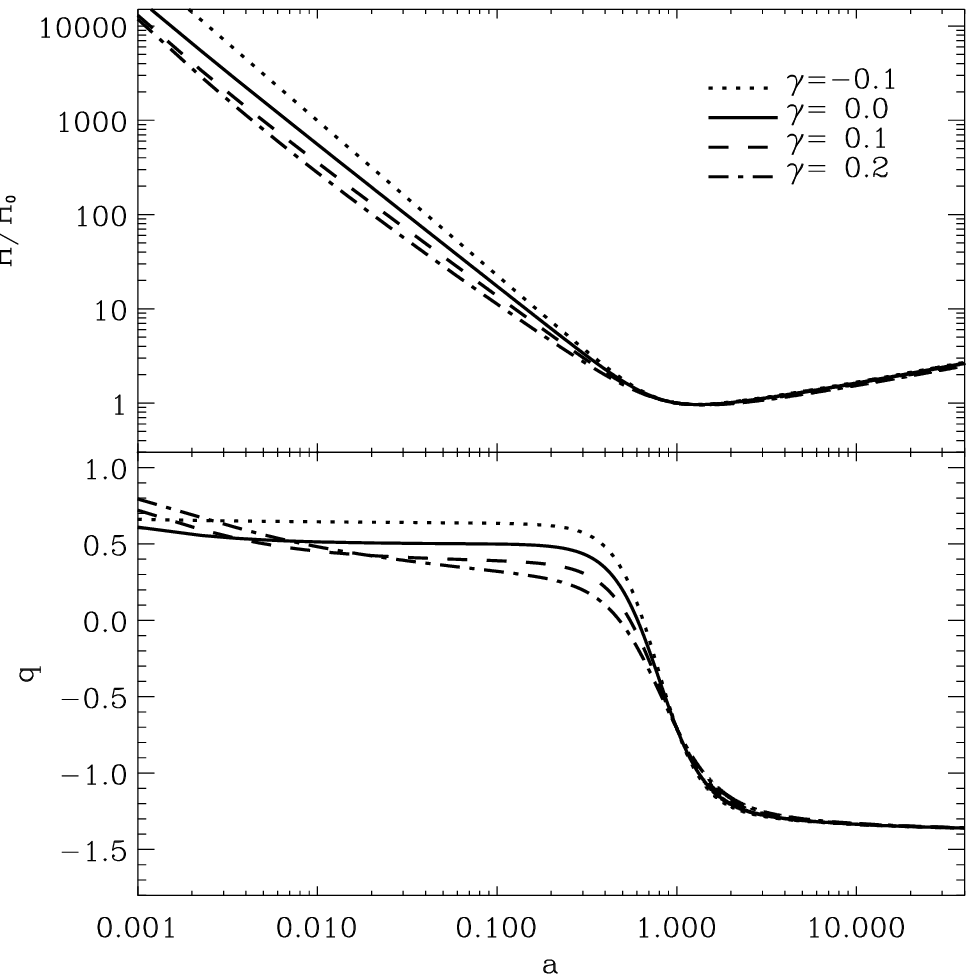}
\caption{The evolution of Hubble parameter $H$ and deceleration parameter $q$ at the different values of interaction parameter $\gamma$ (the dotted line - $\gamma =-0.1$, the solid line - $\gamma =0$, the dashed line - $\gamma =0.1$, the dash-dotted line - $\gamma =0.2$): a - the quintessence dark energy, b - the phantom one. The parameters of models are the same as in the fig. \ref{fig:rcr1}}
\label{fig:dm2}
\end{figure}

In the fig. \ref{fig:dm1} the behaviour of $w$ and $\rho_{de}$ for quintessence dark energy and phantom is shown for the same values of parameters as in the fig. \ref{fig:rcr1} and for the values of interaction parameter $\gamma=-0.1,\,0.0,\,+0.1,\,+0.2$. For the positive values of $\gamma$ the quantity $\rho_{de}$ is always positive. For the quintessence dark energy it is a monotonically decreasing function which asymptotically approaches to the constant value when $a\rightarrow\infty$, and for the phantom one it decreases at first to some minimal value, and then again increases to $+\infty$ when $a\rightarrow\infty$. For the negative value $\gamma=-0.1$ for both quintessence and phantom dark energies $\rho_{de}$ increases from $-\infty$ to 0 and higher, it is the negative at first, and in the process of expansion becomes positive. When changing the sign of $\rho_{de}$ on the opposite, as we see from the figures, $w$ has the singularity of second kind.

Now, having the expressions for energy densities of components, the Friedmann equations can be used for the analysis of impact of such model of interaction between the dark components on the dynamics of expansion of the Universe. We take following the expressions for energy densities of baryonic and relativistic matters: $\rho_{b}=\rho_{b}^{(0)}a^{-3}$, $\rho_{r}=\rho_{r}^{(0)}a^{-4}$.

In the fig. \ref{fig:dm2} the evolution of  $H/H_{0}$ and $q$ is shown for the same values of parameters, as in the fig. \ref{fig:rcr1}. For the interaction parameters $\gamma=-0.1,\,0.0,\,+0.1,\,+0.2$ one can see that for the quintessence dark energy $H/H_{0}\rightarrow\infty$, $q\rightarrow1$ when $a\rightarrow0$, and
$$H/H_{0}\rightarrow(-\Omega_{de}\frac{w_{0}-c_{a}^{2}}{1+c_{a}^{2}})^{1/2}, \ \ \  q\rightarrow-1$$ when $a\rightarrow\infty$. For the phantom dark energy $H/H_{0}\rightarrow\infty$, $q\rightarrow1$ when $a\rightarrow0$ and $H/H_{0}\rightarrow\infty$, $q\rightarrow\frac{1}{2}+\frac{3}{2}c_{a}^{2}$ when $a\rightarrow\infty$. Under above conditions of positivity of $\rho_{de}$, the quantity $H^{2}$ is positive always. One can see, that this is true for the values of interaction parameter $\gamma=0.0,\,+0.1,\,+0.2$. For $\gamma=-0.1$, despite of that $\rho_{de}$ becomes the negative, the total density of components stay always the positive, therefore the quantity $H^{2}$ is always positive, and $q$ smoothly goes from +1 in the early Universe to -1 in the future in the case of quintessence dark energy, or $(1+3c_a^2)/2$ in the case of phantom.

\section*{Conclusions}

The dynamics of expansion of the Universe in the cosmological model with dynamical dark energy, which interacts with dark matter gravitationally and non-gravitationally was analysed. The three types of interaction, which cause the energy-momentum exchange between them, - independent on densities of components, proportional to the density of dark energy and proportional to the density of dark matter were considered. For all cases we have obtained the analytical dependencies $w(a)$, $\rho_{de}(a)$ and $\rho_{dm}(a)$, which are the exact solutions of the conservation equations of energy for the dark components. It was shown, that in all cases the dependencies of energy densities on $a$ are the smooth function, which at the certain values of parameters of dark energy and the interaction parameter can take the negative values. For each case the range of values of parameters, at which the density of dark components are positive for an arbitrary $a$ was found. Common for all cases is the condition of positivity of 
interaction parameter. In the representation (\ref{J0})-(\ref{Jdm}) this means, that only in the case of interaction in which the density of dark matter is always positive is the flow of energy from the dark energy to the dark matter, and the density of dark energy is positive, if and only if the value of interaction parameter not exceed some quantity, found for the each model. The other common feature of such models are the non-zero asymptotic values of densities of dark components when $a\rightarrow\infty$, if the dark energy is the quintessence. If the dark energy is phantom one, then the asymptotic value of its density when $a\rightarrow0$ is  constant for all acceptable values of interaction parameter in the models (\ref{J0}) and (\ref{Jde}). In the model (\ref{Jdm}), when the energy flow is proportional to the density of dark matter, there are possible the variants of special behaviour of the phantom dark energy: $\rho_{de}\rightarrow\infty$ when $a\rightarrow0$ and $a\rightarrow\infty$, or $\rho_{de}
=const$ when $w_0,\,c_a^2<-1$ in the models (\ref{J0}) and (\ref{Jde}). The interaction, which causes the flow of energy from the dark matter to the dark energy, always leads to fast decrease of density of dark matter to zero and transition to the negative values, which we consider as the non-physical solution. The fig. \ref{fig:rcr1}, \ref{fig:de1}, \ref{fig:dm1} confirm these conclusions.

The type of interaction and the value of interaction parameter between the dark components, as seen from the fig. \ref{fig:rcr2}, \ref{fig:de2} and \ref{fig:dm2}, have influence on the dynamics of expansion of Universe - the Hubble parameter and the deceleration parameter, which can be used for identifying the type and the strength of interaction, or at least the upper limits of values of interaction parameter.

The work was supported by the project of Ministry of Education and Science of Ukraine "The dark components and evolutionary stages of formation of high-scale structure of Universe, galaxies, stars and supernova remnants" (state registration number 0113U003059).


\begin{thebibliography}{}
\bibitem{Abdalla2013} Abdalla E., Ferreira E. G. M., Quintin J., Wang B. New Evidence for Interacting Dark Energy from BOSS // 2014.-arXiv:1412.2777v2.

\bibitem{Amendola2000} Amendola L. Coupled quintessence // Phys. Rev. D.-2000.-62, 043511.

\bibitem{Amendola2007} Amendola L., Campos G. C., Rosenfeld R. Consequences of dark matter-dark energy interaction on cosmological parameters derived from type Ia supernova data // Phys. Rev. D.-2007.-75, 083506.

\bibitem{Amendola2003} Amendola L., Quercellini C. Tracking and coupled dark energy as seen by the Wilkinson Microwave Anisotropy Probe // Phys. Rev. D.-2003.-68, 023514.

\bibitem{Amendola2010} Amendola L., Tsujikawa S. Dark Energy: Theory and Observations.-Cambridge, England, Cambridge University Press.-2010.-491p.

\bibitem{Bolotin2013}  Bolotin Yu. L., Kostenko A., Lemets O. A., Yerokhin D. A. Cosmological Evolution With Interaction Between Dark Energy And Dark Matter // Int. J. Mod. Phys. D.-2015.-24, 1530007.-132 pages.

\bibitem{Caldera2009} Caldera-Cabral G., Maartens R., Urena-Lopez L. A. Dynamics of interacting dark energy // Phys. Rev. D.-2009.-79, 063518.

\bibitem{Copeland2006} Copeland E. J., Sami M., Tsujikawa S. Dynamics of dark energy // Int. J. Mod. Phys. D.-2006.-15, No. 11.-p. 1753-1936.

\bibitem{Campo2006} del Campo S., Herrera R., Olivares G., Pavon D. Interacting models of soft coincidence // Phys. Rev. D.-2006.-74, 023501.

\bibitem{Elahi2015} Elahi P. J., Lewis G. F., Power C., et al. Hidden from view: Coupled Dark Sector Physics and Small Scales // Mon. Not. Roy. Astr. Soc.-2015.-452.-p.1341-1352.

\bibitem{Goncalves2015} Goncalves R. S., Carvalho G. C., Alcaniz J. S. A low-z test for interacting dark energy // 2015.-arXiv:1507.01921v1.

\bibitem{Gumjudpai2005} Gumjudpai B., Naskar T., Sami M., Tsujikawa S. Coupled dark energy: towards a general description of the dynamics // JCAP.-2005.-06, 007.

\bibitem{Guo2007} Guo Z. K., Ohta N., Tsujikawa S. Probing the coupling between dark components of the universe // Phys. Rev. D.-2007.-76, 023508.

\bibitem{Vacca2009} La Vacca G., Kristiansen J. R., Colombo L. P. L., et al. Do WMAP data favor neutrino mass and a coupling between Cold Dark Matter and Dark Energy? // JCAP.-2009.-04, 007.

\bibitem{Novosyadlyj2013} Novosyadlyj B., Pelykh V., Shtanov Yu., Zhuk A. Dark energy: observational evidence and theoretical models, eds. V. Shulga.-Kyiv, Akademperiodyka.-2013.-380p.

\bibitem{Sergijenko2010} Novosyadlyj B., Sergijenko O. Scalar field models of dark energy with barotropic equation of state: properties and observational constraints from different datasets // Proceedings of the 10th G. Gamow’s Odessa Astronomical Conference-Summer School Astronomy and Beyond: Cosmomicrophysics, Cosmology and Gravitation, Astrophysics, Radio Astronomy and Astrobiology. Одесса, Украина, Астропринт, 2010.-p.12-21.

\bibitem{Novosyadlyj2010} Novosyadlyj B., Sergijenko O., Apunevych S., Pelykh V. Properties and uncertainties of scalar field models of dark energy with barotropic equation of state // Phys. Rev. D.-2010.-82, 103008.

\bibitem{Novosyadlyj2012} Novosyadlyj B., Sergijenko O., Durrer R., Pelykh V. Do the cosmological observational data prefer phantom dark energy? // Phys. Rev. D.-2012.-86, 083008.

\bibitem{Penzo2015} Penzo C., Maccio A. V., Baldi M., et al. Effects of Coupled Dark Energy on the Milky Way and its Satellites // 2015.-arXiv:1504.07243v1.

\bibitem{Pollina2015} Pollina G., Baldi M., Marulli F., Moscardini L. Cosmic voids in coupled dark energy cosmologies: the impact of halo bias // 2015.-arXiv:1506.08831v1.

\bibitem{Pourtsidou2013} Pourtsidou A., Skordis C., Copeland E. J. Models of dark matter coupled to dark energy // Phys. Rev. D.-2013.-88, 083505.

\bibitem{Sergijenko2015} Sergijenko O., Novosyadlyj B. Sound speed of scalar field dark energy: Weak effects and large uncertainties // Phys. Rev. D.-2015.-91, 083007.

\bibitem{Wei2007} Wei H., Zhang S. N. Observational H(z) data and cosmological models // Phys. Lett. B.-2007.-644.-p. 7-15.

\bibitem{Zimdahl2001} Zimdahl W., Pavon D., Chimento L. P. Interacting quintessence // Phys. Lett. B.-2001.-521.-p. 133-138.

\end{thebibliography}
\end{document}